\documentclass[prb,twocolumn]{revtex4-2}

\usepackage{amsmath,amssymb,physics,dsfont}
\usepackage{color}
\usepackage{hyperref}
\usepackage{microtype}
\usepackage{braket}
\usepackage{epstopdf}
\usepackage{graphicx} 
\usepackage{dcolumn}
\usepackage{bm}
\usepackage{amsmath}
\usepackage{epsfig}
\usepackage{float}
\usepackage{epstopdf}
\begin{document}

\title{Vortex-Controlled Quasiparticle Multiplication and Self-Growth Dynamics in Superconducting Resonators
	}

\author {Joong M. Park$^{1}$, Martin Mootz$^{1}$, Richard H. J. Kim$^{1}$, Zhixiang Chong$^{1,2}$, Samuel Haeuser $^{1,2}$, Randall K. Chan$^{1,2}$, Liang Luo$^{1}$,   
 Dominic P. Goronzy$^{3}$, Mark C. Hersam$^{3,4,5}$, Ilias E. Perakis$^{6}$, Akshay A Murthy$^{7}$, Alexander Romanenko$^{7}$, Anna Grassellino$^{7}$, and Jigang Wang$^{1,2}$}

\affiliation{$^1$Ames National Laboratory - U.S. DOE, Ames, Iowa 50011, U.S.A.} 
\affiliation{$^2$Department of Physics and Astronomy, Iowa State University, Ames, Iowa 50011, U.S.A.}
\affiliation{$^3$Department of Materials Science and Engineering, Northwestern University, Evanston, IL 60208, U.S.A.}
\affiliation{$^4$Department of Chemistry, Northwestern University, Evanston, IL 60208, U.S.A.}
\affiliation{$^5$Department of Electrical and Computer Engineering, Northwestern University, Evanston, IL 60208, U.S.A.}
\affiliation{$^{6}$ Department of Physics, University of Alabama at Birmingham, Birmingham, Alabama 35294-1170, U.S.A}
\affiliation{$^7$Fermi National Accelerator Laboratory, Batavia, Illinois 60510, U.S.A.}

\date{\today}

\begin{abstract}
Even in the quantum limit, non-equilibrium quasiparticle (QP) populations induce QP poisoning that irreversibly relaxes the quantum state and significantly degrades the coherence of transmon qubits. 
A particularly detrimental yet previously unexplored mechanism arises from QP multiplication facilitated by vortex trapping in superconducting quantum circuits, where a high-energy QP relaxes by breaking additional Cooper pairs and amplifying the QP population due to the locally reduced excitation gap and enhanced quantum confinement within the vortex core.
Here we directly resolve this elusive QP multiplication process by revealing vortex-controlled QP self-generation in a highly nonequilibrium regime preceding the phonon bottleneck of QP relaxation.
At sufficiently low fluence, femtosecond-resolved magneto-reflection spectroscopy directly reveals a continuously increasing QP population that is strongly dependent on magnetic-field-tuned vortex density and absent at higher excitation fluences.
Quantitative analysis of the emergent QP pre-bottleneck dynamics further reveals that, although the phonon population saturates within $\simeq$10~ps, both free and trapped QPs continue to grow in a self-sustained manner--hallmarks of the long-anticipated QP-vortex interactions in nonequilibrium superconductivity.
We estimate a substantial increase of $\sim$34\% in QP density at vortex densities of $\sim$ 100 magnetic flux quanta per $\mathrm{\mu m^{2}}$. 
Our findings establish a powerful spectroscopic tool for uncovering QP multiplication and reveal vortex-assisted QP relaxation as a critical materials bottleneck whose mitigation will be essential for resolving QP poisoning and enhancing coherence in superconducting qubits.

\end{abstract}

\maketitle
\section{Introduction}
Carrier multiplication constitutes a unifying mechanism across diverse quantum materials, encompassing multiple-exciton generation in quantum dots, impact ionization in semiconductors, and carrier-carrier multiplication in photovoltaic absorbers~\cite{CM_1, CM_2}. The superconducting analogue--QP multiplication--represents a strongly nonequilibrium relaxation channel detrimental to maintaining coherence, wherein a single high-energy QP decays by breaking additional Cooper pairs and generating multiple lower-energy QPs that collectively amplify the QP population even at the millikelvin temperatures relevant to qubit operation.
QP multiplication is expected to be particularly efficient near vortex cores, where the locally suppressed superconducting gap and strong quantum confinement create a unique nanoscale environment that amplifies electronic interactions and promotes QP generation without involving hot phonons. To date, however, no experiment has directly resolved vortex-dependent QP multiplication and its self-sustained dynamics.
The central challenge is achieving experimental access to the femtosecond-resolved QP trapping, relaxation, and phonon-coupling dynamics that govern nonequilibrium superconducting vortex states. Understanding these processes is vital for mitigating QP poisoning and thereby advancing coherence in superconducting circuits~\cite{QPL1,QPL2}.

Under nonequilibrium conditions, the conventional understanding of light-induced QP kinetics in superconductors~\cite{hot_1, hot_2, hot_3, hot_4} and microelectronic devices~\cite{natc2021} has focused primarily on their interactions with hot phonons. As illustrated in Fig.~1(a), a key process is the phonon bottleneck effect, where QP recombination into Cooper pairs  generates high-energy phonons that, in turn, break additional Cooper pairs. This feedback loop delays net recombination and sustains nonequilibrium QP populations over extended timescales. The  Rothwarf--Taylor (R-T) model has captured this phonon bottleneck regime of nonequilibrium QP dynamics and hot phonons in superconductors~\cite{RT1967,RT1974,RTScalapino}. 
Yet, the {\em pre-bottleneck regime}--preceding the onset of the quasi-equilibrium state--remains far less understood. This early regime is characterized by highly nonthermal QP distributions~\cite{QC_1}, quantum coherence~\cite{QC_2, QC_3, QC_4, QC_5}, and transient many-body dynamics~\cite{hot_2} that can  critically influence the subsequent relaxation pathways~\cite{KabanovPRL2005,YBCOKabanov}. In particular, direct interactions between QPs and vortices can trigger QP multiplication even without changes in the phonon population, underscoring the need to advance both experiment and theory on vortex-state dynamics beyond the conventional hot-phonon framework of superconducting nonequilibrium kinetics.

\begin{figure*}[!tbp]
	\begin{center}
		\includegraphics[scale=.5]{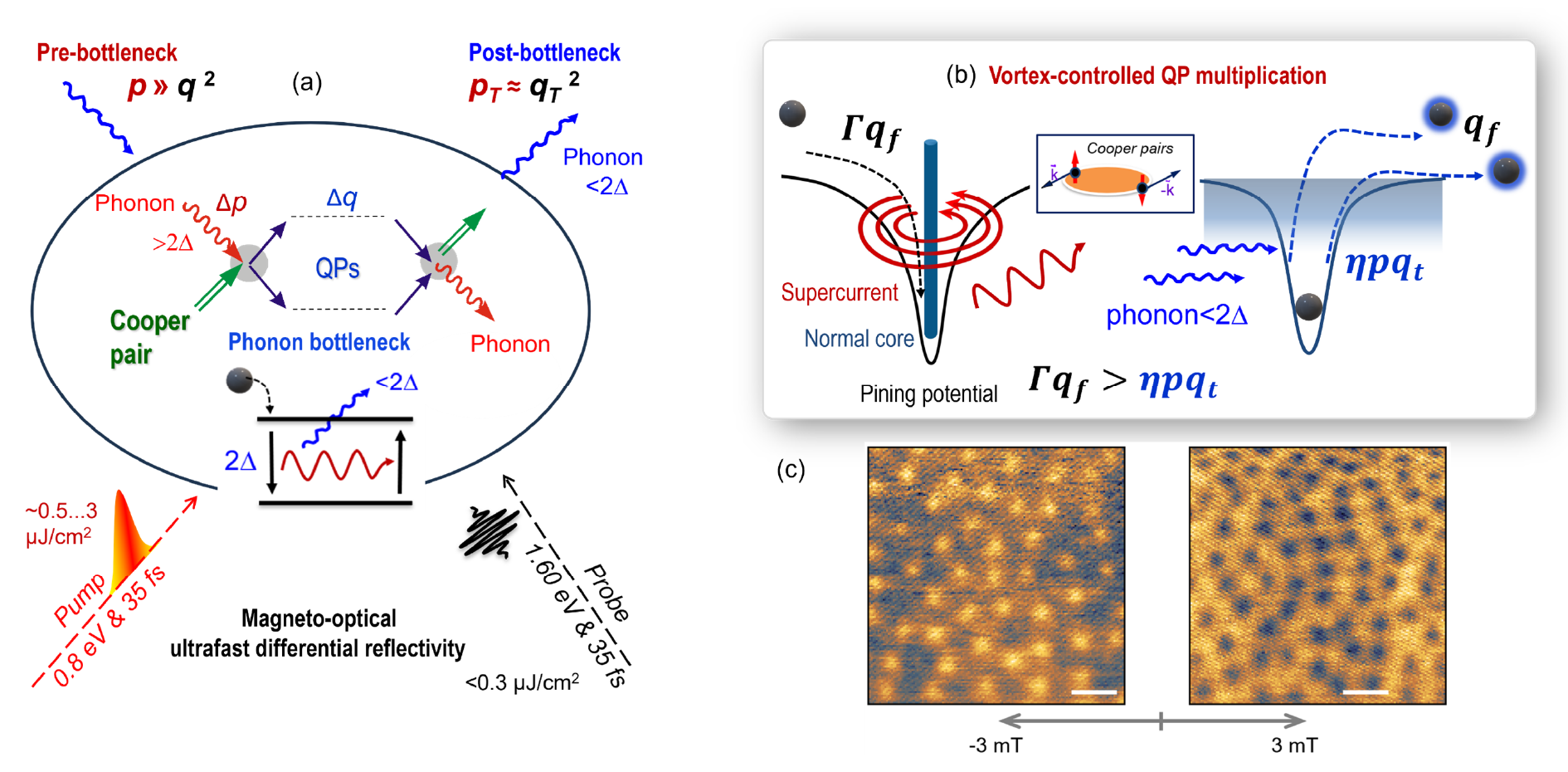}
		\caption{(a) Schematic illustration of quasiparticle (QP) generation and recombination processes in the nonequilibrium superconducting (SC) state without vortices. Here, $q$ denotes the QP density and $p$ denotes the phonon density; $q_T$ and $p_T$ indicate their respective thermal equilibrium values. 	
		(b) Schematic illustration of quasiparticle trapping by a single magnetic vortex in the SC state. The dynamics are governed by the detailed balance among trapped QPs ($q_t$), free QPs ($q_f$), and phonon population ($p$) (main text). Here, $\Gamma$ denotes the vortex-induced QP trapping rate, while $\eta$ represents the QP detrapping rate.
		(c) Magnetic force microscopy (MFM) image of vortices in a Nb thin film under $\pm3$~mT magnetic fields. Each spot corresponds to a single magnetic vortex. Scale bar is 1~$\mu$m.}
        \label{fig1} 
	\end{center}
\end{figure*}

A particularly compelling, yet under-explored, aspect of the QP dynamics arises in the {\em mixed state} of type-II superconductors, where magnetic flux penetrates the material as quantized vortices. As illustrated in Fig.~1(b), each vortex comprises of a normal state core--a region where the SC gap vanishes--embedded in a surrounding SC matrix. These cores serve as nanoscale traps for QPs and host low-energy bound states, acting effectively as artificial quantum dots supporting discrete energy levels within the SC gap~\cite{APL1998,PRL2014}.
This produces a highly inhomogeneous landscape for energy relaxation, QP localization, and dynamical regeneration. However, the hallmark signatures of vortex-mediated QP dynamics have so far remained unobserved, primarily because of the lack of measurement access to femtosecond-scale, highly nonequilibrium and nonthermal QP responses in vortex states of superconducting systems.

Concurrently, in superconducting quantum devices, \emph{two-level systems} (TLSs) and their fluctuation dominate microwave dissipation at sub-kelvin temperatures and MHz frequencies~\cite{TLSloss1,TLSloss2}, whereas QP-induced losses, i.~e., \emph{qubit poisoning}, emerge as the principal decoherence mechanism above 1~K~\cite{QPpoison} and can proceed to 100s of GHz and THz frequency~\cite{hot_1, hot_2}. These losses, largely insensitive to microwave drive power, have also been observed at lower temperatures through background-radiation-induced QP poisoning~\cite{COS_1, COS_2, COS_3, COS_4} revealing an additional dissipation pathway in superconducting qubits~\cite{QPlowT,QPrf}. However, the microscopic origins of nonequilibrium QP generation, and in particular the contribution of magnetic-vortex trapping and de-trapping dynamics, remain to be explored~\cite{vortex, fermi_1, fermi_2}. Although powerful techniques are available to visualize vortex configurations in real space~\cite{kim_1} and microwave frequency~\cite{fermi_1, fermi_2}, achieving the resolution required to probe QP-vortex interactions on femtosecond timescales and THz frequency dynamics is crucial for revealing how vortex states influence nonequilibrium QP generation and dynamics.


In this work, we identify and characterize a \emph{vortex-dominated, pre-bottleneck} regime that hosts an unexpected mechanism of QP multiplication and self-sustained growth under low excitation. Using femtosecond-resolved, low-fluence magneto-pump--probe spectroscopy on superconducting niobium (Nb) resonators under tunable magnetic fields, we resolve ultrafast, magnetic field-dependent QP relaxation dynamics mediated by QP trapping and Cooper-pair breaking within vortex cores. This nonthermal regime sustains a prolonged nonequilibrium state, in which QP multiplication persists even long after the phonon population saturates, revealing a previously unrecognized magnetic-dependent loss channel in superconducting resonators. At higher excitation fluences, this regime collapses into conventional hot phonon dynamics of QP relaxation that are largely magnetic-field-insensitive, reflecting enhanced scattering and thermal activation. Quantitative analysis based on an extended Rothwarf--Taylor model for vortex states shows that, while phonon populations saturate within approximately 10~ps, both free and trapped QP populations continue to grow in a self-sustained manner through QP--vortex scattering, leading to a $\sim$34\% increase in QP density at vortex densities of about 100 magnetic flux quanta per~$\mu\mathrm{m}^{2}$. These results establish magnetic-field-tunable QP multiplication as a new pathway for QP--related loss in nonequilibrium vortex states of superconductors, underscoring the need to mitigate magnetic vortices to overcome QP poisoning in superconducting quantum devices.

\section{Materials and methods}
Nb films were deposited on double-side polished, intrinsic Si(100) wafers with a greater than 10~$k \Omega \cdot \mathrm{cm}$ resistivity.
Prior to deposition, wafers were prepared with a RCA clean that consisted of SC-1 (5:1:1 DI water: ammonium hydroxide: hydrogen peroxide) for 10 min at 80 \textdegree C, then BOE (5:1 ammonium fluoride: hydrofluoric acid) for 2 min, and then SC-2 (5:1:1 DI water: hydrochloric acid: hydrogen peroxide) for 10 min at 80 \textdegree C. Immediately prior to deposition each wafer received another 2 min BOE~(5:1) etch and was then subsequently loaded into the deposition system as quickly as possible to minimize ambient exposure. Nb deposition was done by DC sputtering utilizing a 99.998~\% purity 6” Nb target at 400 W and 3 mTorr process pressure. A thickness of $170\pm 5$~nm was grown at a nominal rate of 5 \AA/s. The base pressure of the deposition system is less than $1 \times 10^{-9}$ Torr. Wafers were coated with S1805 resist for dicing and were cut into 7.5 $\times$ 7.5 $\text{mm}^2$ dies. Dies were subsequently cleaned in NMP for 10 min at 70 \textdegree C, then sonicated in fresh NMP for 10 min and rinsed in IPA and DI water. Dies received a 2 min BOE~(5:1) etch prior to characterization to remove built up Nb oxide from fabrication processing.
The same film was used to fabricate microwave resonator structures to characterize the quality factor and TLS losses, following Ref.~\cite{NbBOE}.

Femtosecond-resolved pump--probe spectroscopy was performed using a fiber-laser amplifier producing 35~fs pulses at a 40~MHz repetition rate. The experimental setup follows previously reported configurations~\cite{Nbmaterials,NbTEM}. As illustrated in Fig.~1(a), the pump beam has a wavelength of 1550~nm (0.8~eV), while the probe beam, generated via second-harmonic generation, has a wavelength of 775~nm (1.6~eV). The pump and probe beams were combined using a dichroic beam combiner and focused in a collinear geometry with a 4-inch focal-length lens. Samples were mounted facing the top window of a dry cryostat equipped with a split-coil superconducting magnet. The differential reflectivity signal, $\Delta R / R$, was detected using a silicon balanced photodiode and a lock-in amplifier with 40~kHz pump modulation. The temporal delay between the pump and probe pulses was controlled by a motorized linear translation stage in the pump beam path. A 900~nm short-pass filter was placed before the detector to block scattered pump light. The beam diameters at the focus were approximately 100~$\mu$m. The pump fluence ranged from 0.5~to~3~$\mu$J/cm$^{2}$, while the probe fluence was kept below~0.3~$\mu$J/cm$^{2}$, ensuring operation well within the low-excitation regime.

\section{Experimental Results}


Figure~\ref{fig1}(a) schematically illustrates the fundamental processes governing conventional nonequilibrium QP generation and relaxation in a niobium superconductor. These dynamics are described by rate equations coupling multiple reservoirs~\cite{RT1967,RT1974}, which capture the interdependent evolution of the QP density $q$ and the high-energy phonon density $p$. In this framework, phonons with energy $\hbar\omega \ge 2\Delta$ can break Cooper pairs to create two QPs, while QPs can recombine to emit phonons with energies $\ge 2\Delta$. At thermal equilibrium, the densities obey the relation $p_T \propto q_T^{2}$. 
{\em High-frequency} photoexcitation perturbs the QP--phonon balance, generating hot phonons that efficiently break additional Cooper pairs. This process re-establishes a feedback loop in which QP recombination produces phonons with energies above the superconducting gap, which are subsequently reabsorbed to create new QPs. The resulting self-sustaining cycle gives rise to a long-lived nonequilibrium state, as the phonon bottleneck inhibits relaxation and extends both QP and phonon lifetimes.
Note that low-energy phonons with energies below $2\Delta$ do not contribute to Cooper-pair breaking or to the phonon bottleneck effect, but instead decay away without further interaction.

The conventional nonequilibrium QP dynamics above is significantly altered by the presence of magnetic vortices, which divides the QP population into trapped QP ($q_t$) and free QP ($q_f$) components. Vortices naturally form under external magnetic fields or microwave driving in type-II superconducting films used in transmon quantum circuits.
Figure~\ref{fig1}(b) illustrates QP trapping by a single magnetic vortex. 
Near magnetic vortices, the superconducting order parameter is strongly suppressed, locally reducing the energy gap to zero within the nanometer-scale vortex cores and enabling QP localization and multiplication. These trapped QPs are spatially separated from regions where phonon--mediated recombination dominates and are confined close around the vortex cores, thereby enhancing QP--QP interactions while diminishing conventional QP--phonon relaxation pathways. Note that the low-energy phonons below $2\Delta$, generated during QP trapping ($q_t$), can still be reabsorbed and assist in Cooper-pair breaking without contributing to the overall phonon population or energy relaxation (blue wiggled lines).

Figure~\ref{fig1}(c) presents a magnetic force microscopy (MFM) image of the real-space vortex distribution in the Nb thin-film superconducting resonators under a weak applied field of $\pm3$~mT. Each bright or dark contrast spot corresponds to a single vortex core with opposite local magnetic-field polarity. The contrast inversion between positive and negative fields arises from the attractive or repulsive interaction between the magnetic MFM tip and the local vortex field. The typical vortex-core diameter (blue column), on the order of tens of nanometers, is comparable to the superconducting coherence length in Nb, underscoring the ability of these nanoscale regions to act as efficient QP traps that can support the proposed QP multiplication process. Furthermore, these vortex states provide a tunable platform for probing and controlling QP--vortex coupling pathways, as elaborated in the following sections.


\begin{figure}[!tbp]
	\begin{center}
		\includegraphics[scale=.6]{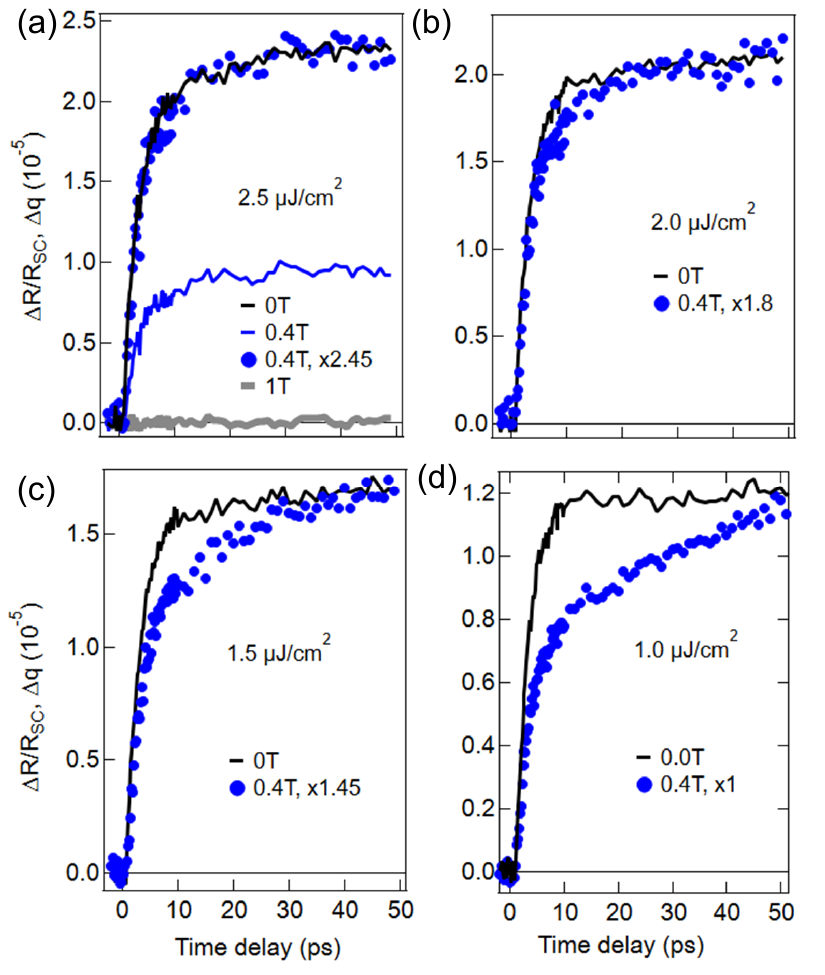}
		\caption{Quasiparticle dynamics in the superconducting state of Nb under 0~T and 0.4~T magnetic fields, measured via the differential reflectivity signal $\Delta R / R_\text{SC}$ at various pump fluences: (a) 2.5~$\mu$J$/$cm$^2$; (b) 2~$\mu$J$/$cm$^2$; (c) 1.5~$\mu$J$/$cm$^2$; (d) 1~$\mu$J$/$cm$^2$. All measurements were done at temperature $T = 2.27$~K. All the $\Delta R / R_\text{SC}$ data measured at 0.4~T (blue circles) are scaled to the 0~T result at 50~ps for easier comparison of dynamics. The blue trace in (a) shows the unscaled raw data measured at 2.5~$\mu$J/cm$^2$, while the gray trace in (a) corresponds to the raw data taken under a magnetic field of 1~T which show $\Delta R / R_\text{SC}$ signals vanish after the complete quenching of superconductivity.
        }
    \label{fig2} 
	\end{center}

\end{figure}	


To establish a baseline result for fs QP dynamics, we first  examine the QP dynamics in the absence of a  magnetic field. Figure~\ref{fig2} (black lines) shows the corresponding differential reflectivity $\Delta R / R_\text{SC}$ associated with QP dynamics in the SC state without magnetic field for four different pump fluences 1, 1.5, 2 and 2.5~$\mu$J$/$cm$^2$. The signal exhibits a sharp rise within $\sim$10~ps, followed by a slow decay exceeding 300~ps. This temporal profile is consistent with conventional phonon bottleneck dynamics involving phonon-limited QP recombination and pair breaking. 
Note that the formation time of the phonon-bottleneck regime, or the pre-bottleneck dynamics, progressively shortens with decreasing pump fluence, consistent with the diminishing hot-phonon population at low excitation.
In comparison, the blue dots in Figs.~\ref{fig2}(a)-(d) present the $\Delta R / R_\text{SC}$ signal measured under external magnetic field of 0.4~T for the four different pump fluences (blue circles). 
Unlike for the two higher fluences with similar decay profiles (Figs.~\ref{fig2}(a)-(b)), the two lower fluences in Figs.~\ref{fig2}(c)-(d) show a pronounced change of the temporal profile when the magnetic field is increased from 0~T to 0.4~T. In particular, we observe a much slower signal rise at low fluences due to the magnetic field. This magnetic-field-dependent buildup of nonequilibrium QP density is distinct from any previously observed ultrafast dynamics in superconducting states, highlighting the salient role of vortex states in nonequilibrium superconductivity.  
At an even higher magnetic field of 1~T, the superconducting signals vanish [gray trace in Fig.~\ref{fig2}(a)], consistent with the magnetic-field-induced suppression of superconductivity. This finding further confirms that the $\Delta R / R_\mathrm{SC}$ response arises exclusively from changes in the condensate density.
The $\Delta R / R_\text{SC}$ signals gradually diminish and vanish near 9~K, consistent with the measured SC transition temperature of $T_\text{c} = 9.1$~K (data not shown). 


\begin{figure}[!tbp]
	\begin{center}
		\includegraphics[scale=.5]{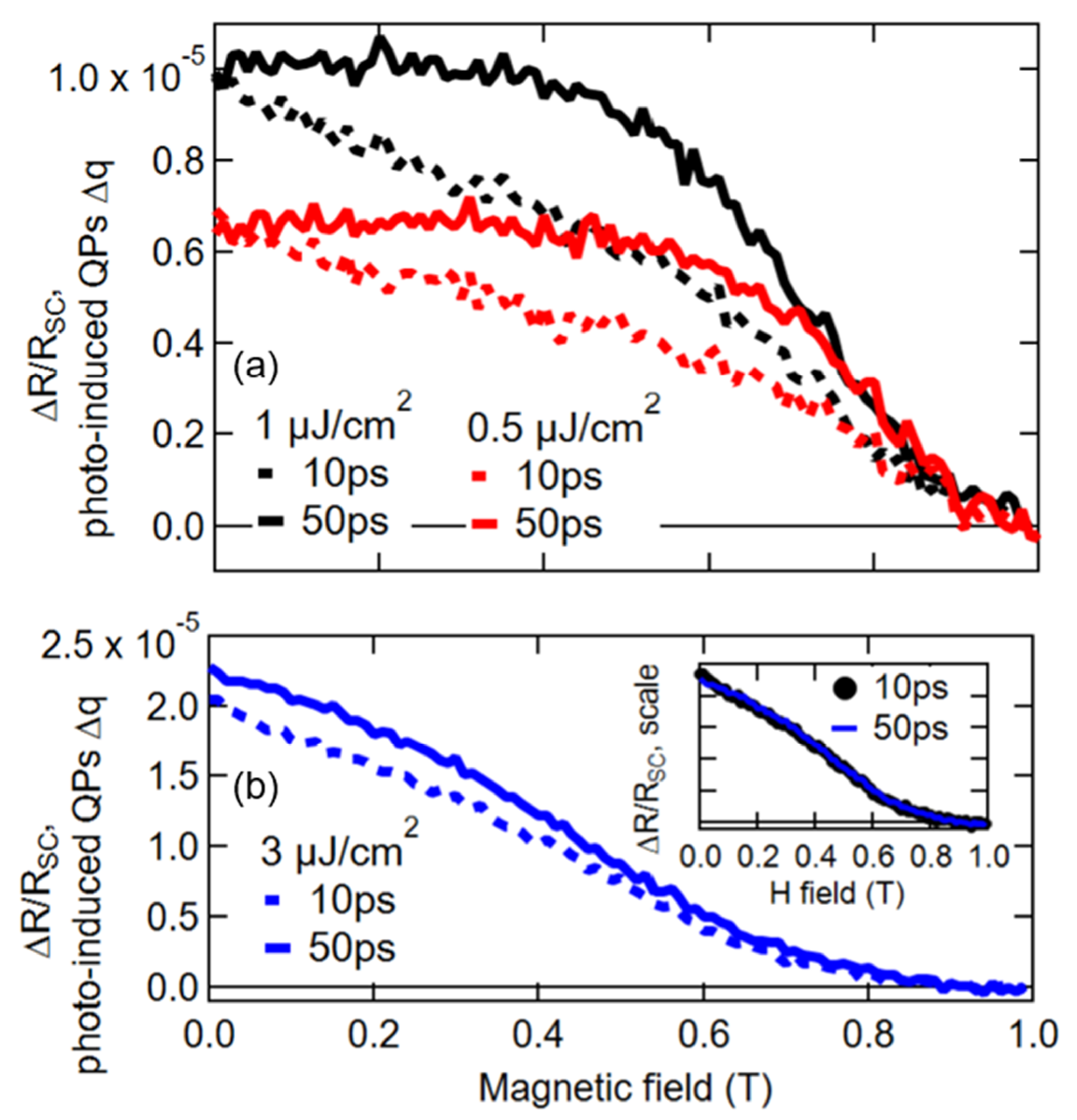}
		\caption{Magnetic-field-dependent, photo-induced superconducting $\Delta R / R_\text{SC}$ components measure quasiparticle change at fixed delay times of 10~ps (dotted lines) and 50~ps (solid lines). Data were recorded at $T = 2.27$~K for pump fluences of 0.5 (red) and 1~$\mu$J/cm$^2$ (black) in (a) and 3~$\mu$J/cm$^2$ in (b). The inset in (b) shows the scaled data for the two time delays in (b).} 
		\label{fig3} 
	\end{center}
\end{figure}	

To elucidate the influence of magnetic vortices on QP formation dynamics, Fig.~\ref{fig3} presents the magnetic--field dependence of $\Delta R / R_\mathrm{SC}$ snapshots at two fixed pump--probe delays: 10~ps (dotted lines) and 50~ps (solid lines).
All measurements were performed at the cryostat base temperature of 2.27~K, with an external magnetic field applied to tune the vortex density.
The magnetic-field dependence exhibits a strong sensitivity to excitation strength, as shown for three representative pump fluences.
The low-fluence regime in Fig.~\ref{fig3}(a) (0.5 and 1~$\mu$J$/$cm$^2$) reveals a pronounced magnetic field dependence for the QP density snapshots at 10~ps and 50~ps. 
For example, at $B = 0.5~\mathrm{T}$, the signals at time delays of 10~ps and 50~ps differ substantially, showing a continuous increase in the QP population during this interval and demonstrating clear magnetic-field-controlled QP multiplication and growth dynamics. This finding is indicative of enhanced QP trapping and localization by vortices, which leads to a spatially-dependent suppression of QP recombination pathways within the vortices that dynamically generate additional QPs as illustrated in Fig.~\ref{fig1}(b). 
In stark contrast to low fluences, the high fluence regime shown in Fig.~\ref{fig3}(b) (3~$\mu$J$/$cm$^2$) exhibits minimal dependence on time delays for magnetic field traces, other than  an overall reduction in amplitude. When normalized, the signals at different magnetic fields (inset) largely overlap, whereas the low--fluence $\Delta R / R_\mathrm{SC}$ snapshots clearly do not scale with one another. These results indicate that, for high fluences, the early-time QP dynamics up to 10~ps is governed primarily by phonon-mediated pair breaking and QP recombination as described by the conventional R-T model in the spatially uniform SC state, with negligible vortex effects and no QP multiplication.
These results provide direct experimental evidence for a field-tunable, pre-bottleneck QP multiplication pathway that emerges uniquely at low excitation densities--one not captured by conventional R-T dynamics without vortices--and disappears at slightly higher fluences, where hot phonons and excess QPs suppress the collective processes due to fragile nonthermal behavior and shallow trapping potentials.
In addition, these observations call for a new model in which vortex--assisted QP trapping enables sustained self-regeneration without hot phonons, introducing a new mechanism for controlling relaxation in superconducting quantum devices.


\begin{figure}[!tbp]
	\begin{center}
        \includegraphics[scale=.4]{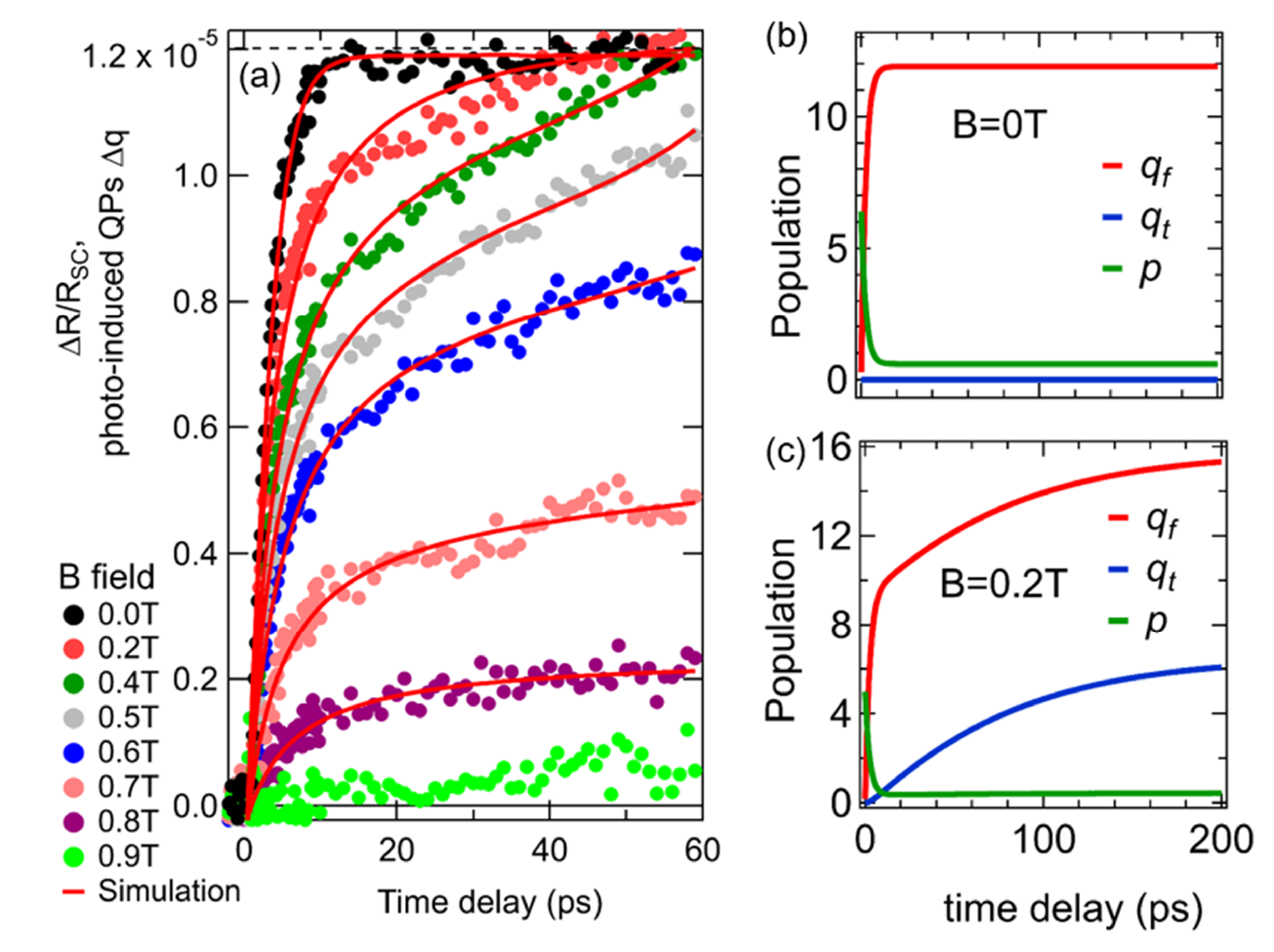}
    \caption{
(a) Comparison between measured superconducting reflectivity transients $\Delta R / R_\text{SC}$ (dots) and fits (solid lines) based on the extended Rothwarf--Taylor model in the vortex states at various magnetic field strengths. The model captures the evolution of QP relaxation dynamics from phonon-bottleneck-dominated decay at zero field to vortex-controlled pre-bottleneck relaxation and QP multiplication at higher fields.
(b) and (c): Model simulations showing the population dynamics of free quasiparticles ($q_f$, red), trapped quasiparticles ($q_t$, blue), and phonons ($p$, green) at 0~T and 0.2~T.}      
		\label{fig4} 
	\end{center}
\end{figure}

\section{Discussion}

To quantitatively model the field-dependent SC QP dynamics observed in Fig.~\ref{fig2}, we  extend the R-T model  by incorporating  QP trapping by magnetic vortices and QP  detrapping by  phonon-driven processes.  In particular, we propose a {\em  three-fluid model} that extends the conventional R-T framework by introducing a third population, trapped QPs ($q_t$), in addition to the conventional free QPs ($q_f$) and high-energy phonons ($p$). The dynamics of these three coupled populations is governed by the following  rate equations:
\begin{align}
&\frac{\mathrm{d}q_f}{\mathrm{d}t} = -R\,q_f^2 + \beta \,p - \Gamma \,q_f + \eta\, p\, q_t\,, \nonumber\\
&\frac{\mathrm{d}q_t}{\mathrm{d}t} = \Gamma\, q_f - \eta\, p\, q_t\,, \nonumber \\
&\frac{\mathrm{d}p}{\mathrm{d}t} = \frac{1}{2} R\, q_f^2 - \frac{1}{2} \beta\, p - \eta\, p\, q_t + \Gamma\, q_t\,. 
\label{eq:RT}
\end{align}
Here, $R$  denotes the QP recombination rate, $\beta$  the phonon-induced pair-breaking rate, $\Gamma$  the vortex-induced QP trapping rate, and $\eta$ the phonon-mediated detrapping rate, including contributions from low-energy phonons below the SC excitation gap. This extended model captures both the conventional phonon bottleneck behavior and the new dissipation channel introduced by QP trapping in the presence of magnetic vortices.

Figure~\ref{fig4}(a) compares the experimental differential reflectivity signals  $\Delta R / R_\text{SC}$ (dots) with numerical fits obtained  using the above  three--fluid model (solid lines) across a range of magnetic fields. The excellent agreement  validates  the model's ability to consistently replicate the key features of the QP multiplication and relaxation dynamics across all magnetic fields. At low fields, the $\Delta R / R_\text{SC}$ signal exhibits a rapid rise  over $\approx 10$~ps followed by saturation--consistent with a buildup of photoinduced QPs that establishes a  phonon bottleneck limiting pair breaking and recombination.  As the  field increases, the relaxation slows and deviates significantly  from the low-field behavior, particularly at later times, reflecting the growing influence of vortex-induced QP trapping and QP multiplication that suppresses the phonon bottleneck leading to the rapid saturation of the signal at zero field. Intriguingly, the extracted population dynamics of free QPs ($q_f$, red), trapped QPs ($q_t$, blue), and phonons ($p$, green) show distinct behaviors at 0~T (Fig.~\ref{fig4}(b)) and 0.2~T (Fig.~\ref{fig4}(c)). The continuous growth of the QP densities for both $q_t$ and $q_f$ is clearly visible in the 0.2~T traces, even after the phonon population ceases to change at approximately 10~ps. These QP multiplication processes are absent in the 0~T traces, indicative of the vortex-controlled QP dynamics.

\begin{figure}[!tbp]
	\begin{center}
        \includegraphics[scale=.4]{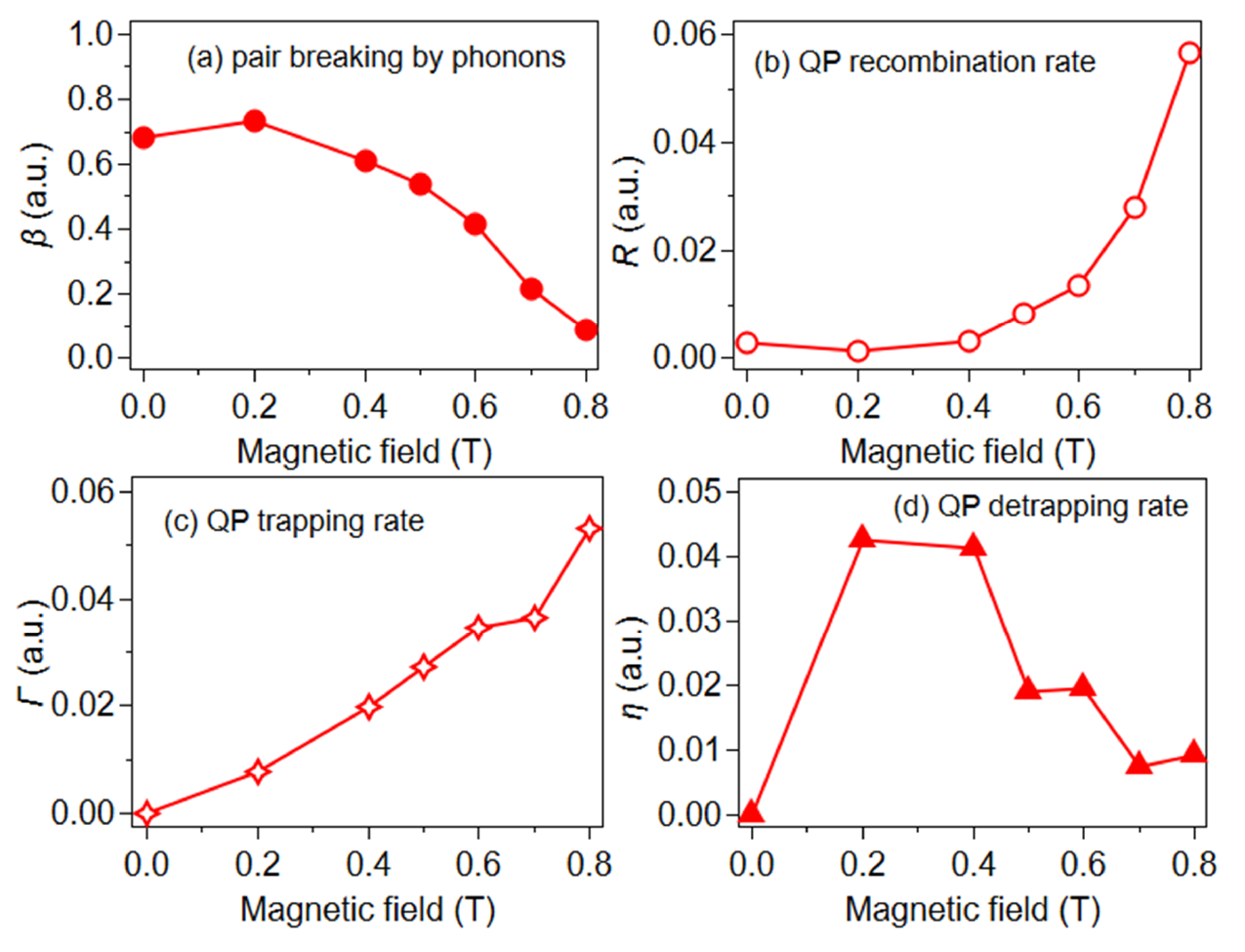}
    \caption{
Magnetic field dependence of the four characteristic rates extracted from the model fits using the extended R-T model in the vortex states:
(a) Pair-breaking rate $\beta$ shows a decrease as the field approaches the critical field $H_\text{c}$, consistent with SC gap suppression and reduced pair-breaking efficiency.
(b) QP recombination rate $R$ increases monotonically with field, reflecting enhanced recombination in vortex-core regions where the SC order parameter is suppressed.
(c) Vortex-induced QP trapping rate $\Gamma$ increases with field, in line with the rising vortex density in the type-II SC state.
(d) Phonon-mediated detrapping rate $\eta$ remains approximately constant up to $H \approx 0.4$~T, then declines near $H_\text{c}$, due to phonon softening and reduced detrapping efficiency.
Together, these trends reveal a crossover from phonon-dominated to vortex-dominated relaxation dynamics as magnetic field increases.}       
		\label{fig5} 
	\end{center}
\end{figure}

To extract physical insight, Fig.~\ref{fig5} shows the  magnetic field dependence of the four  key rates entering our model: $\beta$, $R$, $\Gamma$, and $\eta$. The field dependence of these parameters reflects the evolving change in the balance between  pair-breaking, recombination, vortex trapping, and phonon-mediated detrapping as the magnetic field modulates the SC state and emergence of QP multiplication. We emphasize four key features. First, the phonon--induced pair-breaking rate $\beta$ (Fig.~\ref{fig5}(a)) displays a non-monotonic field dependence. A modest increase at low fields is likely due to partial softening of the SC gap, which enhances the phase space for phonon absorption. However, as the field approaches the upper critical field $H_\text{c}$, $\beta$ decreases sharply--consistent with the  collapse of the SC and  phonon spectrum modifications (e.~g., softening or damping), which suppress phonon efficiency in breaking Cooper pairs. Second, the QP recombination rate $R$ (Fig.~\ref{fig5}(b)) increases with magnetic field. This field enhancement  is attributed to enhanced  QP recombination near the vortex cores, where the SC order parameter is suppressed and the local density of states is broadened.  As the vortex density  increases,  these QP recombination-enhancing regions expand, promoting faster QP annihilation. Third, the QP trapping rate $\Gamma$ (Fig.~\ref{fig5}(c))  increases with magnetic field, consistent with the expected rise in vortex density in the type-II SC state. Each additional vortex core acts as a new potential trap for mobile QPs,  thereby increasing the frequency of trapping events. Fourth, the phonon-mediated detrapping rate $\eta$ (Fig.~\ref{fig5}(d)) remains approximately constant  for fields up to  0.4~T, but decreases near $H_\text{c}$. This drop may  arise  from field-induced changes in the phonon availability and characteristics,  particularly a reduction in the population of phonons with sufficient energy to liberate trapped QPs, or enhanced phonon damping, which limit the system's ability to detrap QPs. 

Taken together, the above trends establish that magnetic vortices fundamentally reshape QP dynamics in low-fluence regimes, by introducing a field-tunable dissipation pathway not affected by the phonon bottleneck. This QP multiplication regime supports a self-sustained QP population via vortex-assisted trapping and phonon-driven pair breaking, even when phonon bottleneck effects are nominally expected to dominate. Using this quantitative analysis in Fig.~\ref{fig4} together with the magnetic-field-dependent vortex density, one can estimate a self-sustained growth of QP density leading to an increase of approximately 34\% at vortex densities of about 100 magnetic flux quanta per~$\mu\mathrm{m}^{2}$. In addition, note that the present measurements were performed at magnetic fields down to approximately 10--100~G (Fig.~\ref{fig3}(a)) to ensure sufficient signal sensitivity for resolving femtosecond QP dynamics, higher than previous studies that probed microsecond--scale fluctuations responsible for microwave loss under fields below 1~G~\cite{fermi_1,fermi_2}. Nevertheless, the underlying QP-vortex interaction revealed here remains directly applicable in the low-trapped-flux regime (below 1~G) characteristic of superconducting quantum devices, where such interactions are expected to play a significant role in determining the energy--relaxation time ($T_1$) and coherence of superconducting qubits.

Critically, the impact of this vortex-mediated mechanism diminishes at high pump fluences, where QP dynamics revert to conventional Rothwarf-Taylor behavior--largely insensitive to magnetic field. This delineates two distinct dynamical regimes: one governed by conventional QP--phonon kinetics, and another defined by emergent vortex-controlled nonequilibrium physics, accessible only under low-excitation conditions.
These insights not only elucidate a previously uncharacteristic component of superconducting relaxation dynamics, but also highlight a promising avenue for coherence engineering in quantum devices. By tuning the magnetic field and excitation fluence, it becomes possible to control QP trapping and dissipation, offering a materials-level strategy for mitigating QP poisoning in superconducting qubits.



\section{Conclusion} 

In summary, we identify and characterize a distinct, magnetic-field-tunable relaxation regime in superconducting Nb, where quasiparticle dynamics evolve from phonon-dominated to vortex-mediated behavior as the excitation fluence is reduced.
Using femtosecond magneto–pump--probe spectroscopy and an extended three-fluid Rothwarf--Taylor framework, we extract the microscopic evolution of scattering and recombination rates, revealing how magnetic vortices open new dissipation channels by trapping quasiparticles and reshaping pair-breaking kinetics.

This crossover demonstrates that external magnetic fields not only modulate quasiparticle populations but fundamentally transform the interactions that determine their lifetimes. The three-fluid model provides a predictive, time-resolved framework for controlling quasiparticle dynamics through simultaneous tuning of magnetic field and excitation strength.

These findings establish vortex-mediated quasiparticle multiplication as a key nonequilibrium loss mechanism and highlight vortex control as a viable route to mitigate quasiparticle poisoning and enhance coherence in superconducting quantum devices. More broadly, this work introduces a new paradigm for probing and engineering ultrafast nonequilibrium processes in topologically structured superconductors, enabling future advances in light--vortex interactions and dissipation-aware quantum materials design~\cite{2D}.

\begin{acknowledgments} 
The sample preparation and pump-probe spectroscopy experiment were supported by the U.S. Department of Energy, Office of Science, National Quantum Information Science Research Centers, Superconducting Quantum Materials and Systems Center (SQMS) under contract No. DE-AC02-07CH11359. 
The simulation was supported by the U.S. Department of Energy, Office of Basic Energy Science, Division of Materials Sciences and Engineering. The Ames Laboratory is operated for the U.S. Department of Energy by Iowa State University under Contract No. DE-AC02-07CH11358. We thank the Rigetti fabrication team for their support in process development and for fabricating the specimens used in the initial stage of this study.
This work was also authored by Fermi Forward Discovery Group, LLC, under Contract No. 89243024CSC000002 with the U.S. Department of Energy, Office of Science, Office of High Energy Physics.
\end{acknowledgments}

\bibliography{ref}

@article{2D,
  author        = {Huang, Chuankun and Mootz, Martin and Luo, Liang and Perakis, Ilias E. and Wang, Jigang},
  title         = {Unlocking Quantum Control and Multi--Order Correlations via Terahertz {2D} Coherent Spectroscopy},
  journal       = {Nat. Rev. Phys.},
  year          = {2025},
  note          = {in press},
  eprint        = {2507.02116},
  archivePrefix = {arXiv},
  primaryClass  = {cond-mat.supr-con},
  url           = {https://arxiv.org/abs/2507.02116}
}

@Article{COS_1,
author={Veps{\"a}l{\"a}inen, Antti P.
and Karamlou, Amir H.
and Orrell, John L.
and Dogra, Akshunna S.
and Loer, Ben
and Vasconcelos, Francisca
and Kim, David K.
and Melville, Alexander J.
and Niedzielski, Bethany M.
and Yoder, Jonilyn L.
and Gustavsson, Simon
and Formaggio, Joseph A.
and VanDevender, Brent A.
and Oliver, William D.},
title={Impact of ionizing radiation on superconducting qubit coherence},
journal={Nature},
year={2020},
volume={584},
pages={551-556},
doi={10.1038/s41586-020-2619-8}
}

@Article{COS_2,
author={Wilen, C. D.
and Abdullah, S.
and Kurinsky, N. A.
and Stanford, C.
and Cardani, L.
and D'Imperio, G.
and Tomei, C.
and Faoro, L.
and Ioffe, L. B.
and Liu, C. H.
and Opremcak, A.
and Christensen, B. G.
and DuBois, J. L.
and McDermott, R.},
title={Correlated charge noise and relaxation errors in superconducting qubits},
journal={Nature},
year={2021},
volume={594},
pages={369-373},
doi={10.1038/s41586-021-03557-5}
}

@Article{COS_3,
author={Cardani, L.
and Valenti, F.
and Casali, N.
and Catelani, G.
and Charpentier, T.
and Clemenza, M.
and Colantoni, I.
and Cruciani, A.
and D'Imperio, G.
and Gironi, L.
and Gr{\"u}nhaupt, L.
and Gusenkova, D.
and Henriques, F.
and Lagoin, M.
and Martinez, M.
and Pettinari, G.
and Rusconi, C.
and Sander, O.
and Tomei, C.
and Ustinov, A. V.
and Weber, M.
and Wernsdorfer, W.
and Vignati, M.
and Pirro, S.
and Pop, I. M.},
title={Reducing the impact of radioactivity on quantum circuits in a deep-underground facility},
journal={Nat. Commun.},
year={2021},
volume={12},
pages={2733},
doi={10.1038/s41467-021-23032-z},
}

@Article{COS_4,
author={McEwen, Matt
and Faoro, Lara
and Arya, Kunal
and Dunsworth, Andrew
and Huang, Trent
and Kim, Seon
and Burkett, Brian
and Fowler, Austin
and Arute, Frank
and Bardin, Joseph C.
and Bengtsson, Andreas
and Bilmes, Alexander
and Buckley, Bob B.
and Bushnell, Nicholas
and Chen, Zijun
and Collins, Roberto
and Demura, Sean
and Derk, Alan R.
and Erickson, Catherine
and Giustina, Marissa
and Harrington, Sean D.
and Hong, Sabrina
and Jeffrey, Evan
and Kelly, Julian
and Klimov, Paul V.
and Kostritsa, Fedor
and Laptev, Pavel
and Locharla, Aditya
and Mi, Xiao
and Miao, Kevin C.
and Montazeri, Shirin
and Mutus, Josh
and Naaman, Ofer
and Neeley, Matthew
and Neill, Charles
and Opremcak, Alex
and Quintana, Chris
and Redd, Nicholas
and Roushan, Pedram
and Sank, Daniel
and Satzinger, Kevin J.
and Shvarts, Vladimir
and White, Theodore
and Yao, Z. Jamie
and Yeh, Ping
and Yoo, Juhwan
and Chen, Yu
and Smelyanskiy, Vadim
and Martinis, John M.
and Neven, Hartmut
and Megrant, Anthony
and Ioffe, Lev
and Barends, Rami},
title={Resolving catastrophic error bursts from cosmic rays in large arrays of superconducting qubits},
journal={Nat. Phys.},
year={2022},
volume={18},
pages={107-111},
doi={10.1038/s41567-021-01432-8}
}

@article{QC_2,
author = {Chuankun Huang  and Martin Mootz  and Liang Luo  and Di Cheng  and Avinash Khatri  and Joong-Mok Park  and Richard H. J. Kim  and Yihua Qiang  and Victor L. Quito  and Yongxin Yao  and Peter P. Orth  and Ilias E. Perakis  and Jigang Wang },
title = {Discovery of an unconventional quantum echo by interference of {Higgs} coherence},
journal = {Sci. Adv.},
volume = {11},
pages = {eads8740},
year = {2025},
doi = {10.1126/sciadv.ads8740}
}

@Article{QC_1,
author={Yang, X.
and Vaswani, C.
and Sundahl, C.
and Mootz, M.
and Gagel, P.
and Luo, L.
and Kang, J. H.
and Orth, P. P.
and Perakis, I. E.
and Eom, C. B.
and Wang, J.},
title={Terahertz-light quantum tuning of a metastable emergent phase hidden by superconductivity},
journal={Nat. Mater.},
year={2018},
volume={17},
pages={586-591},
doi={10.1038/s41563-018-0096-3}
}

@article{QC_3,
  title = {Quantum Coherence Tomography of Light-Controlled Superconductivity},
  author = {Luo, L. and Mootz, M. and Kang, J. H. and Huang, C. and Eom, K. and Lee, J. W. and Vaswani, C. and Collantes, Y. G. and Hellstrom, E. E. and Perakis, I. E. and Eom, C. B. and Wang, J.},
  year = {2023},
  journal = {Nat. Phys.},
  volume = {19},
  pages = {201--209},
  doi       = {10.1038/s41567-022-01827-1}
}

@Article{QC_4,
author={Vaswani, C.
and Kang, J. H.
and Mootz, M.
and Luo, L.
and Yang, X.
and Sundahl, C.
and Cheng, D.
and Huang, C.
and Kim, R. H. J.
and Liu, Z.
and Collantes, Y. G.
and Hellstrom, E. E.
and Perakis, I. E.
and Eom, C. B.
and Wang, J.},
title={Light quantum control of persisting {Higgs} modes in iron-based superconductors},
journal={Nat. Commun.},
year={2021},
volume={12},
pages={258},
doi={10.1038/s41467-020-20350-6}
}

@article{QC_5,
  title={Lightwave-driven superconductivity and forbidden quantum beats by terahertz symmetry breaking},
  author={Yang, X and Vaswani, C and Sundahl, C and Mootz, M and Luo, L and Kang, JH and Perakis, IE and Eom, CB and Wang, J},
  journal={Nat. Photonics},
  volume={13},
  pages={707--713},
  year={2019},
doi={10.1038/s41566-019-0470-y}
}

@article{CM_1,
  title = {High Efficiency Carrier Multiplication in {PbSe} Nanocrystals: Implications for Solar Energy Conversion},
  author = {Schaller, R. D. and Klimov, V. I.},
  journal = {Phys. Rev. Lett.},
  volume = {92},
  pages = {186601},
  year = {2004},
  doi = {10.1103/PhysRevLett.92.186601}
}

@article{CM_2,
author = {Beard, Matthew C. and Knutsen, Kelly P. and Yu, Pingrong and Luther, Joseph M. and Song, Qing and Metzger, Wyatt K. and Ellingson, Randy J. and Nozik, Arthur J.},
title = {Multiple Exciton Generation in Colloidal Silicon Nanocrystals},
journal = {Nano Lett.},
volume = {7},
pages = {2506-2512},
year = {2007},
doi = {10.1021/nl071486l}
}

@article{kim_1,
  title={A sub-2 {Kelvin} cryogenic magneto-terahertz scattering-type scanning near-field optical microscope {(cm-THz-sSNOM)}},
  author={Kim, Richard H J and Park, JM and Haeuser, Samuel J and Luo, Liang and Wang, Jigang},
  journal={Rev. Sci. Instrum.},
  volume={94},
  pages={043702},
  year={2023},
doi = {10.1063/5.0130680},
note      = {Editor's Pick}
}

@misc{fermi_1,
      title={Demonstrating magnetic field robustness and reducing temporal {T1} noise in transmon qubits through magnetic field engineering}, 
      author={Bektur Abdisatarov and Tanay Roy and Daniel Bafia and Roman Pilipenko and Matthew Julian Dubiel and David van Zanten and Shaojiang Zhu and Mustafa Bal and Grigory Eremeev and Hani Elsayed-Ali and Akshay Murty and Alexander Romanenko and Anna Grassellino},
      year={2025},
      eprint={2506.02187},
      archivePrefix={arXiv},
      primaryClass={quant-ph},
      url={https://arxiv.org/abs/2506.02187}, 
}

@article{fermi_2,
    author = {Bafia, D. and Abdisatarov, B. and Pilipenko, R. and Lu, Y. and Eremeev, G. and Romanenko, A. and Grassellino, A.},
    title = {Quantifying trapped magnetic vortex losses in niobium resonators at {mK} temperatures},
    journal = {Appl. Phys. Lett.},
    volume = {127},
    pages = {152601},
    year = {2025},
    doi = {10.1063/5.0282159}
}

@article{QPL1,
  title = {Mitigation of Quasiparticle Loss in Superconducting Qubits by Phonon Scattering},
  author = {Bargerbos, Arno and Splitthoff, Lukas Johannes and Pita-Vidal, Marta and Wesdorp, Jaap J. and Liu, Yu and Krogstrup, Peter and Kouwenhoven, Leo P. and Andersen, Christian Kraglund and Gr\"unhaupt, Lukas},
  journal = {Phys. Rev. Appl.},
  volume = {19},
  pages = {024014},
  year = {2023},
  doi = {10.1103/PhysRevApplied.19.024014}
}

@article{QPL2,
author = {Torres-Castanedo, Carlos G. and Goronzy, Dominic P. and Pham, Thang and McFadden, Anthony and Materise, Nicholas and Masih Das, Paul and Cheng, Matthew and Lebedev, Dmitry and Ribet, Stephanie M. and Walker, Mitchell J. and Garcia-Wetten, David A. and Kopas, Cameron J. and Marshall, Jayss and Lachman, Ella and Zhelev, Nikolay and Sauls, James A. and Mutus, Joshua Y. and McRae, Corey Rae H. and Dravid, Vinayak P. and Bedzyk, Michael J. and Hersam, Mark C.},
title = {Formation and Microwave Losses of Hydrides in Superconducting Niobium Thin Films Resulting from Fluoride Chemical Processing},
journal = {Adv. Funct. Mater.},
volume = {34},
pages = {2401365},
doi = {https://doi.org/10.1002/adfm.202401365},
year = {2024}
}

@article{hot_1,
  title = {Ultrafast nonthermal terahertz electrodynamics and possible quantum energy transfer in the ${\mathrm{Nb}}_{3}\mathrm{Sn}$ superconductor},
  author = {Yang, X. and Zhao, X. and Vaswani, C. and Sundahl, C. and Song, B. and Yao, Y. and Cheng, D. and Liu, Z. and Orth, P. P. and Mootz, M. and Kang, J. H. and Perakis, I. E. and Wang, C.-Z. and Ho, K.-M. and Eom, C. B. and Wang, J.},
  journal = {Phys. Rev. B},
  volume = {99},
  pages = {094504},
  year = {2019},
  doi = {10.1103/PhysRevB.99.094504}
}

@Article{hot_2,
author={Cheng, Bing
and Cheng, Di
and Lee, Kyuho
and Luo, Liang
and Chen, Zhuoyu
and Lee, Yonghun
and Wang, Bai Yang
and Mootz, Martin
and Perakis, Ilias E.
and Shen, Zhi-Xun
and Hwang, Harold Y.
and Wang, Jigang},
title={Evidence for $d$-wave superconductivity of infinite-layer nickelates from low-energy electrodynamics},
  journal   = {Nat. Mater.},
  volume    = {23},
  pages     = {775},
  year      = {2024},
    doi={10.1038/s41563-023-01766-z}
}

@article{hot_3,
author = {B. Q. Song  and X. Yang  and C. Sundahl  and J.-H. Kang  and M. Mootz  and Y. Yao  and I. E. Perakis  and L. Luo  and C. B. Eom  and J. Wang },
title = {Ultrafast Martensitic Phase Transition Driven by Intense Terahertz Pulses},
journal = {Ultrafast Science},
volume = {3},
pages = {0007},
year = {2023},
doi = {10.34133/ultrafastscience.0007}
}

@article{hot_4,
  title = {Nonequilibrium Pair Breaking in $\mathrm{Ba}({\mathrm{Fe}}_{1\ensuremath{-}x}{\mathrm{Co}}_{x}{)}_{2}{\mathrm{As}}_{2}$ Superconductors: Evidence for Formation of a Photoinduced Excitonic State},
  author = {Yang, X. and Luo, L. and Mootz, M. and Patz, A. and Bud'ko, S. L. and Canfield, P. C. and Perakis, I. E. and Wang, J.},
  journal = {Phys. Rev. Lett.},
  volume = {121},
  pages = {267001},
  year = {2018},
  doi = {10.1103/PhysRevLett.121.267001}
}

@article{RT1967,
  title = {Measurement of Recombination Lifetimes in Superconductors},
  author = {Rothwarf, Allen and Taylor, B. N.},
  journal = {Phys. Rev. Lett.},
  volume = {19},
  pages = {27--30},
  year = {1967},
  doi = {10.1103/PhysRevLett.19.27}
}

@article{RT1974,
  title = {Quasiparticle Lifetimes and Microwave Response in Nonequilibrium Superconductors},
  author = {Rothwarf, A. and Sai-Halasz, G. A. and Langenberg, D. N.},
  journal = {Phys. Rev. Lett.},
  volume = {33},
  pages = {212--215},
  year = {1974},
  doi = {10.1103/PhysRevLett.33.212}
}

@article{RTScalapino,
  title = {Kinetic-equation approach to nonequilibrium superconductivity},
  author = {Chang, Jhy-Jiun and Scalapino, D. J.},
  journal = {Phys. Rev. B},
  volume = {15},
  pages = {2651--2670},
  year = {1977},
  doi = {10.1103/PhysRevB.15.2651}
}

@article{KabanovPRL2005,
  title = {Kinetics of a Superconductor Excited with a Femtosecond Optical Pulse},
  author = {Kabanov, V. V. and Demsar, J. and Mihailovic, D.},
  journal = {Phys. Rev. Lett.},
  volume = {95},
  pages = {147002},
  year = {2005},
  doi = {10.1103/PhysRevLett.95.147002}
}

@article{YBCOKabanov,
  title = {Quasiparticle relaxation dynamics in superconductors with different gap structures: Theory and experiments on ${\mathrm{YBa}}_{2}{\mathrm{Cu}}_{3}{\mathrm{O}}_{7\mathrm{\ensuremath{-}}\mathrm{\ensuremath{\delta}}}$},
  author = {Kabanov, V. V. and Demsar, J. and Podobnik, B. and Mihailovic, D.},
  journal = {Phys. Rev. B},
  volume = {59},
  pages = {1497--1506},
  year = {1999},
  doi = {10.1103/PhysRevB.59.1497}
}

@Article{natc2021,
author={Weng, Qianchun
and Yang, Le
and An, Zhenghua
and Chen, Pingping
and Tzalenchuk, Alexander
and Lu, Wei
and Komiyama, Susumu},
title={Quasiadiabatic electron transport in room temperature nanoelectronic devices induced by hot-phonon bottleneck},
journal={Nat. Commun.},
year={2021},
volume={12},
pages={4752},
doi={10.1038/s41467-021-25094-5}
}

@article{APL1998,
    author = {Ullom, J. N. and Fisher, P. A. and Nahum, M.},
    title = {Magnetic field dependence of quasiparticle losses in a superconductor},
    journal = {Appl. Phys. Lett.},
    volume = {73},
    pages = {2494-2496},
    year = {1998},
    doi = {10.1063/1.122493}
}

@article{PRL2014,
  title = {Trapping a Single Vortex and Reducing Quasiparticles in a Superconducting Resonator},
  author = {Nsanzineza, I. and Plourde, B. L. T.},
  journal = {Phys. Rev. Lett.},
  volume = {113},
  pages = {117002},
  year = {2014},
  doi = {10.1103/PhysRevLett.113.117002}
}

@article{TLSloss1,
    author = {Valli{\`e}res, Andre and Russell, Megan E. and You, Xinyuan and Garcia-Wetten, David A. and Goronzy, Dominic P. and Walker, Mitchell J. and Bedzyk, Michael J. and Hersam, Mark C. and Romanenko, Alexander and Lu, Yao and Grassellino, Anna and Koch, Jens and McRae, Corey Rae H.},
    title = {Loss tangent fluctuations due to two-level systems in superconducting microwave resonators},
    journal = {Appl. Phys. Lett.},
    volume = {126},
    pages = {124001},
    year = {2025},
    doi = {10.1063/5.0253375}
}

@article{TLSloss2,
    author = {Sage, Jeremy M. and Bolkhovsky, Vladimir and Oliver, William D. and Turek, Benjamin and Welander, Paul B.},
    title = {Study of loss in superconducting coplanar waveguide resonators},
    journal = {J. Appl. Phys.},
    volume = {109},
    pages = {063915},
    year = {2011},
    doi = {10.1063/1.3552890}
}

@article{QPpoison,
  author    = {Aumentado, Jos{\'e} and Catelani, Gianluigi and Serniak, Kyle},
  title     = {Quasiparticle poisoning in superconducting quantum computers},
  journal   = {Phys. Today},
  volume    = {76},
  number    = {8},
  pages     = {34--39},
  year      = {2023},
  doi       = {10.1063/PT.3.5291}
}

@article{QPlowT,
  title = {Effects of nonequilibrium quasiparticles in a thin-film superconducting microwave resonator under optical illumination},
  author = {Budoyo, R. P. and Hertzberg, J. B. and Ballard, C. J. and Voigt, K. D. and Kim, Z. and Anderson, J. R. and Lobb, C. J. and Wellstood, F. C.},
  journal = {Phys. Rev. B},
  volume = {93},
  pages = {024514},
  year = {2016},
  doi = {10.1103/PhysRevB.93.024514}
}

@article{QPrf,
  title = {Evidence of a Nonequilibrium Distribution of Quasiparticles in the Microwave Response of a Superconducting Aluminum Resonator},
  author = {de Visser, P. J. and Goldie, D. J. and Diener, P. and Withington, S. and Baselmans, J. J. A. and Klapwijk, T. M.},
  journal = {Phys. Rev. Lett.},
  volume = {112},
  pages = {047004},
  year = {2014},
  doi = {10.1103/PhysRevLett.112.047004}
}

@article{vortex,
author={Taupin, M.
and Khaymovich, I. M.
and Meschke, M.
and Mel'nikov, A. S.
and Pekola, J. P.},
title={Tunable quasiparticle trapping in {Meissner} and vortex states of mesoscopic superconductors},
journal={Nat. Commun.},
year={2016},
volume={7},
pages={10977},
doi={10.1038/ncomms10977}
}

@article{NbBOE,
doi = {10.1088/2633-4356/ad88cc},
year = {2024},
volume = {4},
pages = {045101},
author = {Kopas, Cameron J and Goronzy, Dominic P and Pham, Thang and Torres Castanedo, Carlos G and Cheng, Matthew and Cochrane, Rory and Nast, Patrick and Lachman, Ella and Zhelev, Nikolay Z and Vallières, André and Murthy, Akshay A and Oh, Jin-su and Zhou, Lin and Kramer, Matthew J and Cansizoglu, Hilal and Bedzyk, Michael J and Dravid, Vinayak P and Romanenko, Alexander and Grassellino, Anna and Mutus, Josh Y and Hersam, Mark C and Yadavalli, Kameshwar},
title = {Enhanced superconducting qubit performance through ammonium fluoride etch},
journal = {Mater. Quantum Technol.},
}

@article{Nbmaterials,
AUTHOR = {Park, Joong-Mok and Chong, Zhi Xiang and Kim, Richard H. J. and Haeuser, Samuel and Chan, Randy and Murthy, Akshay A. and Kopas, Cameron J. and Marshall, Jayss and Setiawan, Daniel and Lachman, Ella and Mutus, Joshua Y. and Yadavalli, Kameshwar and Grassellino, Anna and Romanenko, Alex and Wang, Jigang},
TITLE = {Probing Non-Equilibrium Pair-Breaking and Quasiparticle Dynamics in {Nb} Superconducting Resonators Under Magnetic Fields},
JOURNAL = {Materials},
VOLUME = {18},
YEAR = {2025},
ARTICLE-NUMBER = {569},
DOI = {10.3390/ma18030569}
}

@article{NbTEM,
title = {Exploring the relationship between deposition method, microstructure, and performance of {Nb}/{Si}-based superconducting coplanar waveguide resonators},
journal = {Acta Mater.},
volume = {276},
pages = {120153},
year = {2024},
doi = {https://doi.org/10.1016/j.actamat.2024.120153},
author = {Jin-Su Oh and Cameron J. Kopas and Jayss Marshall and Xiaotian Fang and Kamal R. Joshi and Amlan Datta and Sunil Ghimire and Joong-Mok Park and Richard Kim and Daniel Setiawan and Ella Lachman and Joshua Y. Mutus and Akshay A. Murthy and Anna Grassellino and Alex Romanenko and John Zasadzinski and Jigang Wang and Ruslan Prozorov and Kameshwar Yadavalli and Matt Kramer and Lin Zhou}
}

\end{document}